# Magnetic compensation phenomenon by Nd doping in 'spin surplus' Sm based ferromagnetic intermetallic SmScGe

P. D. Kulkarni[1], S. K. Dhar[1], P. Manfrinetti[2] and A. K. Grover[1]

[1]Department of Condensed Matter Physics and Materials Science, Tata Institute of Fundamental Research, Colaba, Mumbai-400005.

[2]INFM and Dipartimento di Chimica e Chimica Industriale, Universit " a di Genova, Via Dedecaneso 31, Genova 16146, Italy

Email: prasanna@tifr.res.in

*Abstract*

*The magnetization compensation phenomenon is observed for $Sm_{1-x}Nd_xScGe$ at x=0.09 at a temperature ~90 K in a small applied magnetic field, which establishes the 'spin surplus' status of magnetic moment of Sm in the host matrix. We also noted that the sample profiles in QD SQUID magnetometer become asymmetric in some situations indicating significant contributions from multiple moments higher than the dipole moment. We have successfully accounted for them using a fitting procedure involving the higher order magnetic moments.*

## INTRODUCTION

Zero magnetization state in a ferromagnetic 4f rare earth (RE) intermetallic is typically realized by partially replacing a given RE element by an element belonging to the other half of the RE series. Samarium element belonging to the first half of the RE series has, however, a special status. In free ion ($Sm^{3+}$) form, its magnetic moment is dominated by the contribution from the orbital momentum ($<L_z>$), which exceeds the opposing contributions from the spin moment ($<2S_z>$). In some situations, the crystalline and exchange field induced admixture effects of low lying higher excited states into the ground state can make the spin contribution to the magnetic moment dominates that from the orbital angular momentum. In such a circumstance, the magnetic compensation in a ferromagnetic Sm intermetallic would get accomplished by replacing the Sm ions by RE elements belonging to the first half of the RE series, instead of by the ions belonging to the second half of the RE series. We reckoned that the ferromagnetic intermetallic compound SmScGe ($T_c$ ~270 K) is one such system [1] which is 'spin surplus' as regards the Sm moment. SmScGe has tetragonal CeScSi-type structure, in which Sm ions occupy a unique site. We explored the partial substitution of $Sm^{3+}$ ions by $Gd^{3+}$ and $Nd^{3+}$ ions in this compound, and have indeed found that the magnetic compensation in it gets brought about by the latter ions, and not by that of the former. We present some aspects of the studies in progress.

## EXPERIMENTAL

A polycrystalline sample of the $Sm_{0.91}Nd_{0.09}ScGe$ was prepared by melting together stoichiometric amounts of SmScGe and NdScGe. Magnetic measurements were carried out using a Quantum Design SQUID magnetometer.

## RESULTS AND DISCUSSION

Fig. 1 shows the field cooled cool down (FCC) magnetization as a function of temperature at few selected fields in $Sm_{0.91}Nd_{0.09}ScGe$ sample. Global ordering temperature is observed at ~260 K. Below this temperature, the net magnetization increases and peaks at ~210. Below this temperature, the net magnetization decreases. In a magnetic field of 5 mT, magnetization values are seen to scatter while crossing the zero value. Fig. 2 shows the SQUID profiles at three different temperatures in the temperature range where the magnetization values show a scatter. It can seen that the SQUID profile is nearly symmetric at a temperature ~131 K. However, below this temperature, a larger asymmetry in the profile could be noted. (e.g., see profiles at ~85 K and ~75 K). The regression values in the analysis performed by QD software were also found to decrease from ~99% to ~60% in the said temperature range. The net magnetization values in the interval ~130 K to 70 K in Fig. 1, as given by the analysis of the raw data by SQUID magnetometer software, are not reliable. The same behavior (i.e., large scatter in data) is witnessed for FCC curves recorded at fields other than 5 mT. A careful look at the SQUID profiles across the temperature interval 140 K to 70 K indicated that these profiles suffer large deviations from the usual dipolar profile, expected for a coil array having second derivative configuration. We, therefore, decided to record the SQUID profiles from 140 K to 50 K and reanalyze the data taking into account admixture contributions from higher multipole moments (quadruple and octuple) into that from the dipole moment, as per a prescription due to Guy and Howarth [2]. The reliable value of the net magnetization alongwith the contributions from quadruple and octuple moments get obtained by following a fitting procedure referred to above.

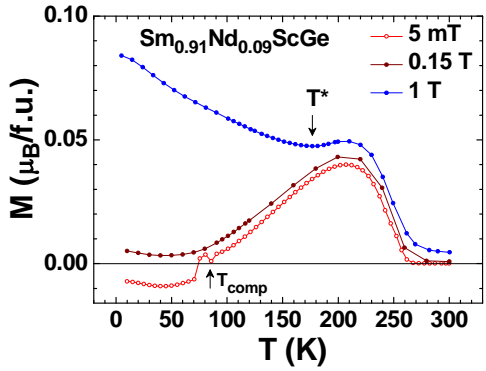

Fig. 1. Field cooled cooldown (FCC) magnetization curves in $Sm_{0.91}Nd_{0.09}ScGe$ at the fields indicated.

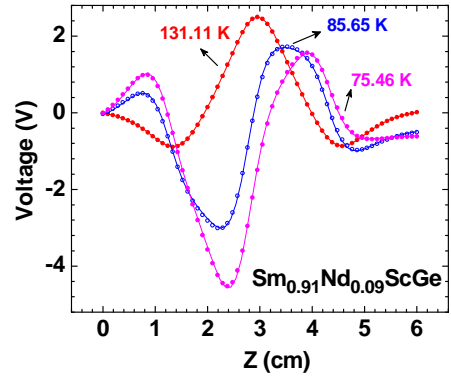

Fig. 2. Flux profiles obtained in Quantum Design SQUID magnetometer at three different temperatures, 135K, 85K and 75 K in $Sm_{0.91}Nd_{0.09}ScGe$.

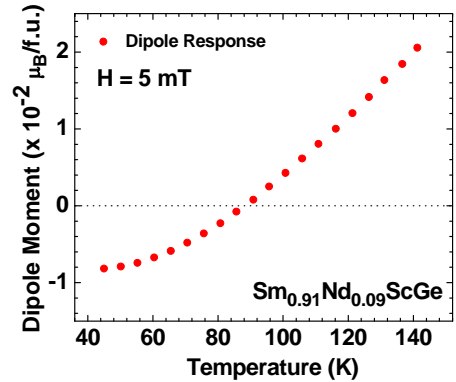

Fig. 3. Extracted values of dipole moment as a function of temperature in H = 5 mT in $Sm_{0.91}Nd_{0.09}ScGe$.

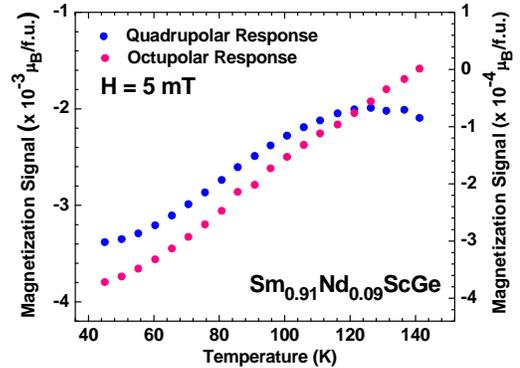

Fig. 4. Extracted values of quadruple and octuple moment obtained, from fitting of the SQUID profile, vs. temperature in $Sm_{0.91}Nd_{0.09}ScGe$.

The continuous lines passing through data points in Fig. 2 are a consequence of such fittings. Fig. 3 shows the magnetization vs. temperature from 140 K to 45 K in a field of 5 mT. Note that magnetization data now smoothly cross across zero magnetization value at ~90 K. Fig. 4 shows the plots of temperature variation of contributions from quadruple and octuple moments in the interval, 140 K to 45 K, these contribution are one or two orders of magnitude smaller than that of the dipole contribution. The FCC magnetization curve in fields of 0.15T and 1T (Fig. 1) in $Sm_{0.91}Nd_{0.09}ScGe$ sample indicate an evolution in behavior corresponding to reorientation of Sm and Nd moments across the temperature, at which the turnaround in magnetization happens. Curious observation is the large field dependence in the reorientation temperature ($T^*$), in this alloy. Such a large variation in $T^*$ values in a given sample is not evident in 'zero magnetization spin ferromagnets' like $Sm_{1-x}Gd_xAl_2$ (x=0.1,0.2,0.3) and $Nd_{0.75}Ho_{0.25}Al_2$ [3,4]. Efforts to record the fingerprint of reorientation behavior in $Sm_{0.91}Nd_{0.09}ScGe$ in the in-phase ac susceptibility data measured in superposed dc magnetic fields did not yield affirmative result. Multidomain structure in magnetically ordered structure is expected to result in significant values of higher multiple moments, (than dipole). Usually their contributions are smaller and they do not interfere with extraction of dipolar contribution in magnetization measurements, even in samples belonging to the 'zero magnetization' category (e.g., $Sm_{1-x}Gd_xAl_2$, $Nd_{0.75}Ho_{0.25}Al_2$, etc.). Studies in different RE based 'zero magnetization' spin ferromagnets may reveal information on the evolution in their domain structures.

## CONCLUSIONS

'Spin surplus' character in the magnetic moment of Sm in SmScGe, is established via observation of magnetic compensation by Nd substituted (9%) of Sm. The higher order contributions are separated from that due to magnetic moment in $Sm_{0.91}Nd_{0.0.9}ScGe$ sample.